# *Double and triple ionization of isocyanic acid*

J.H.D. Eland[1], R.J. Squibb[2], A.J. Sterling[3], M. Wallner[2], A. Hult Roos[2], J. Andersson[2], V. Axelsson[2], E. Johansson[2], A. Teichter[2], S. Stranges[4,5], B. Brunetti[5], J.M. Dyke[6], F. Duarte[3], and R. Feifel[2]

[1]Department of Chemistry, Physical and Theoretical Chemistry Laboratory, Oxford University, South Parks Road, Oxford OX1 3QZ, United Kingdom
[2]Department of Physics, University of Gothenburg, Origovägen 6B, SE-412 96 Gothenburg, Sweden
[3]Chemistry Research Laboratory, University of Oxford, Mansfield Road, Oxford OX1 3TA, United Kingdom
[4]IOM-CNR Tasc, SS-14, Km 163.5 Area Science Park, Basovizza 34149, Trieste, Italy
[5]Dipartimento di Chimica e Tecnologie del Farmaco, Universitá Sapienza, Rome I-00185, Italy
[6]School of Chemistry, University of Southampton, Highfield, Southampton SO17 1BJ, United Kingdom

**Abstract**

Double and triple ionization spectra of isocyanic acid have been measured using multi-electron and ion coincidence techniques combined with synchrotron radiation and compared with high-level theoretical calculations. Vertical double ionization at an energy of 32.8±0.3 eV forms the $^3A''$ ground state in which the $HNCO^{2+}$ ion is long-lived. The vertical triple ionization energy is determined as 65±1 eV. The core-valence double ionization spectra resemble the valence photoelectron spectrum in form, and their main features can be understood on the basis of a simple and rather widely applicable Coulomb model based on the characteristics of the molecular orbitals from which electrons are removed. Characteristics of the most important dissociation channels are examined and discussed.

**Introduction**

Because of the importance of isocyanic acid, HNCO, in terrestrial environments [1-3] and in the interstellar medium [4], spectra of this molecule and its singly charged ion $HNCO^+$ have been widely studied both experimentally and theoretically. A rather complete listing of the earlier spectroscopic work is given in a recent paper by Holzmeier et al. [5] on its normal and resonant Auger spectra. The electronic and geometric structure of neutral HNCO and of its three most stable isomers have been calculated [6,7], as has the structure of the singly positive ion [8,9]. The dynamics of fragmentation of its positive ions both singly and doubly charged have long been studied by mass-spectrometric methods [10,11] and more recently by coincidence methods [8,12]. The states of the singly charged ions, seen in the photoelectron spectrum, could be correlated to the dissociation pathways [8], but this was not possible for the doubly charged ions [12] as no spectroscopic information on them existed at that time. We now report spectra of its double and triple ionization by photon impact, obtained using a multi-electron coincidence technique combined with synchrotron radiation in the soft X-ray region.

The coincidence techniques used in the present work consists in the detection and energy analysis of all the electrons or electrons and ions emitted in encounters between single high energy photons and single gas-phase molecules. It relies mainly on the use of a magnetic-bottle time-of-flight analyser [13] for the electrons, and monochromatic photons from an electron storage ring or laboratory light source for ionization.

**Experimental methods**

Experiments were done at beamline UE52/SGM of the electron storage ring BESSY-II at the Helmholtz Zentrum Berlin when the ring was operated in single-bunch mode. Because the period of 800 ns between bunches in this mode is much shorter than the flight times of low energy electrons (up to 5000 ns) or of ions, a mechanical chopper [14], synchronized to the ring pulses, was used to reduce the inter-pulse period to about 12 μs in the source region of the magnetic bottle [14]. At that point the light pulses intersect an effusive jet of target gas from a hollow needle in the divergent field (ca 1 kG) of a permanent magnet, which directs almost all the emitted photoelectrons towards a distant detector. The 2 m long flight path is surrounded by a solenoid whose field lines guide the electrons to the microchannel plate detector where their arrival times relative to the light pulses are registered. Flight times are converted to electron kinetic energies with the help of calibration using well-known photoelectron and Auger electron energies. The energy resolution is limited mainly by imperfect parallelism of the electron trajectories and for these experiments could be expressed as a numerical ratio E/ΔE ≈ 50. For experiments in the laboratory the same electron spectrometer was used, but the light source was a pulsed discharge in low-pressure He followed by a toroidal-grating monochromator [15], providing 40.8 eV photons from the HeIIα atomic emission line. To examine the dissociations of doubly charged $HNCO^{2+}$ ions the same magnetic bottle was augmented with an in-line time-of-flight mass selector, which has been fully described before [16]. Briefly, a pulsed ion drawout field is applied to the source region once all electrons have escaped into the field-free flight tube. Ions are accelerated towards a microchannel plate detector by fields imposing the time-focussing conditions. Because a less intense divergent magnetic field is used, the electron resolution E/ΔE under these conditions is about 20 while the mass resolution (FWHM) is about 50.

Isocyanic acid was prepared by the reaction between potassium cyanate and an excess of molten stearic acid at 86 C. The reagents were finely ground and scrupulously dried over $P_2O_5$ in vacuo for several days before use. The raw reaction products were condensed in a liquid nitrogen ($LN_2$ ) trap, then repeatedly vacuum-distilled into a trap cooled by a solid $CO_2$-acetone bath before final but short-period storage at $LN_2$ temperature for admission to the apparatus.

**Computational methods**

Calculations were carried out using the ORCA suite of programs (version 4.1.1) [17]. Multi-reference calculations for HNCO, HNCO$^{++}$ and HNCO$^{+++}$ were run at the CASSCF(8,6)/ano-TZVP, CAS(6,6)/ano-TZVP and CAS(5,6)/ano-TZVP levels of theory, respectively[18]. These active spaces incorporate 8 (neutral), 6 (doubly ionized) and 5 (triply ionized) electrons distributed in the bonding, non-bonding and antibonding orbital configurations of both the in-plane and out-of-plane π-systems (1a'' 8a' 9a' 2a''10a'3a''). Structures were confirmed to be local minima by the absence of imaginary frequencies upon calculation of the Hessian. Dynamic correlation was incorporated with the multi-reference configuration interaction (MRCI) method with explicit singles and doubles [19]. DFT calculations of isomers of [H,N,C,O] were carried out at the B3LYP/6-311G(*d*,*p*) level of theory for direct comparison with previous studies [9,20].

**Results and discussion**

Figure 1 shows valence double ionisation spectra produced by photoionization at 40.8 eV and at 90-100 eV photon energy. In the ionization energies of the different bands, the upper spectrum in Fig. 1, closely resembles the well-resolved normal Auger spectra of HNCO recorded by Holzmeier et al. [5], but the relative band intensities are quite different.

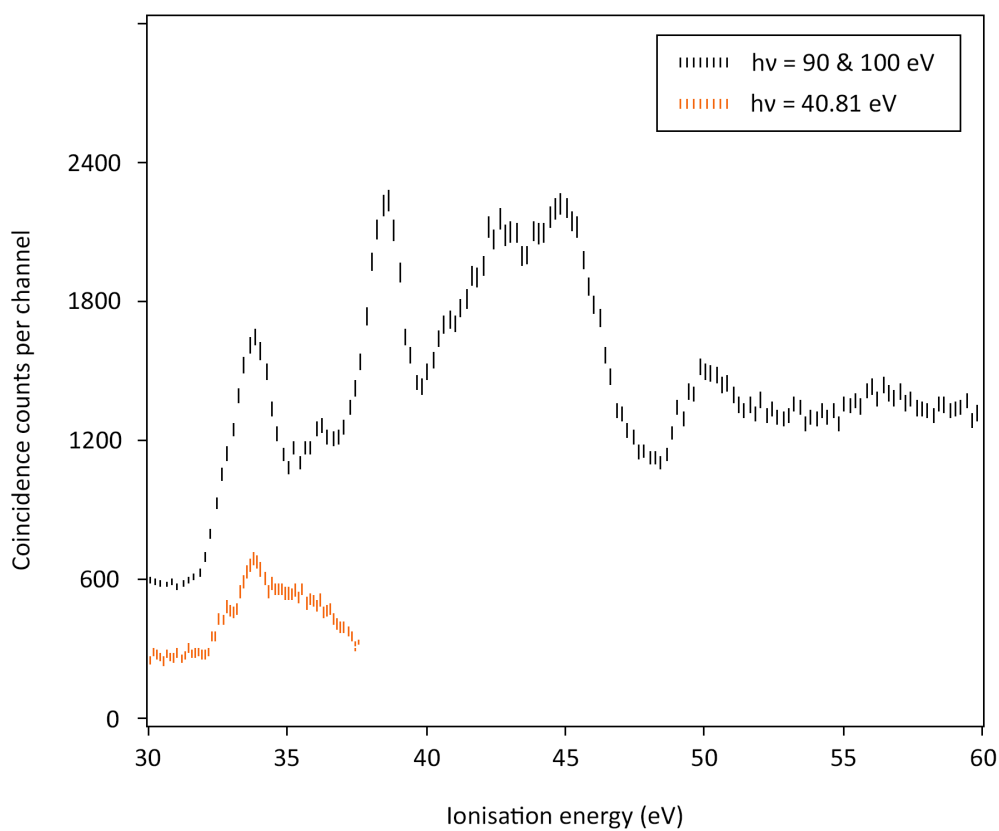

Figure 1. Valence double ionization of HNCO. The lower spectrum was measured in the laboratory at 40.81 eV photon energy. The upper spectrum is a composite made from two independent runs at 100 eV and one run at

90 eV, which all give the same spectrum. Experimental points are represented by 2σ error bars. Resolution is estimated as 0.3 eV in the lower spectrum and 1.3 eV for the first band in the upper spectrum.

The closest resemblance is to the calculated spectrum of singlet doubly ionised states from initial $C1s^{-1}$ ionization [5], a comparison which also demonstrates that each band actually represents a group of states. The states populated by valence photoionization are expected to include triplets as well as singlets, and the peak at 32.8 eV in the better–resolved 40.8 eV spectrum and the one at 36 eV in the 100 eV spectrum, which do not appear in the experimental or calculated Auger spectra [5], may represent triplets. It is relevant to an interpretation of the spectrum to recall the orbital ordering in neutral HNCO (omitting the inner shells) which is

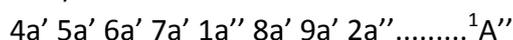
4a' 5a' 6a' 7a' 1a'' 8a' 9a' 2a''.........$^1A''$

where all orbitals are doubly occupied and the two outermost orbitals (2a'' and 9a') are the out-of-plane and in-plane components corresponding to the non-bonding $\pi_g$ orbital of isoelectronic $CO_2$. The next inner pair of orbitals (1a'' and 8a') correspond in the same way to the bonding $\pi_u$ orbital of $CO_2$.

The difference in binding energy between the in-plane and out-of-plane non-bonding orbitals has been calculated as 0.8 eV [5], in agreement with the vertical ionization energy difference seen in the photoelectron spectrum [21]. In a new calculation at the CASSCF/MRCI level of theory the triplet state from $2a''^{-1}9a'^{-1}$ ionization is found to lie 0.7 eV lower in energy than the first singlet, so we conclude that the ground state of the doubly charged $HNCO^{++}$ ion is $^3A''$ from $2a''^{-1}9a'^{-1}$ ionization with onset at 32.2±0.2 eV and peak at 32.8±0.3 eV. At this level of theory the vertical ionization energy is calculated as 32.5 eV, in excellent agreement with experiment (Fig. 2).

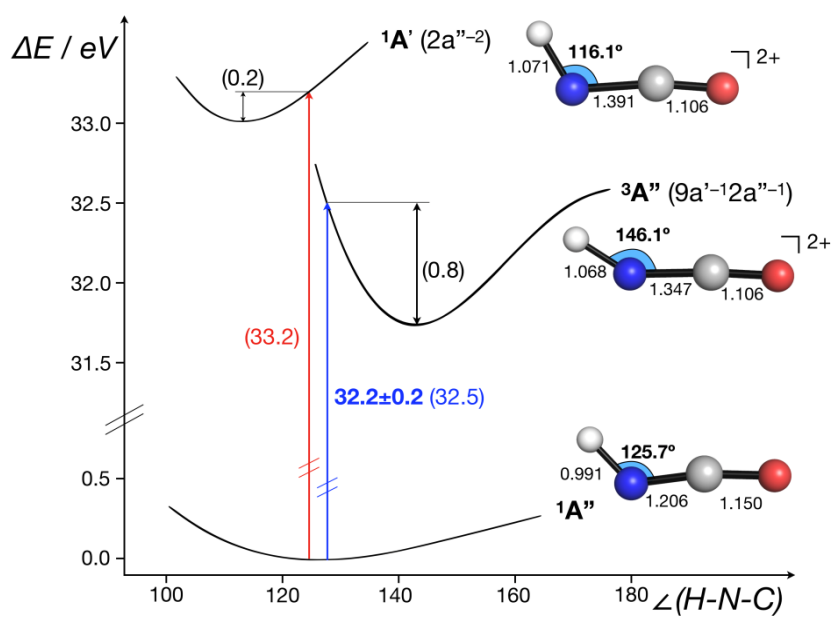

Figure 2. Double ionization energies and key calculated structural data for HNCO (**1A''**) to the lowest energy triplet (**3A''**, $9a'^{-1}2a''^{-1}$) and singlet (**1A'**, $2a''^{-2}$) states. Experimental ionization energy in bold, calculated values

at CASSCF/MRCI in parentheses. Ground state geometry calculated at CASSCF(8,6)/ano-TZVP. Geometries of doubly ionised states calculated at CASSCF(6,6)/ano-TZVP. Potential energy surfaces for the ∠H-N-C angle calculated at the level of theory used for optimisation of equilibrium structures (not to scale).

This triplet ground state is seen as the first peak in the 40.8 eV double ionization spectrum and as a barely perceptible shoulder at the same energy on the low-energy side of the first peak in the 100 eV spectrum. CASSCF/MRCI calculations predict the adiabatic double ionization to be 0.8 eV lower in energy than the vertical, where the removal of a single electron from both the in-plane 9a' and 2a'' orbitals (3A'', Fig. 2) causes the molecule to change from a bent to a more linear equilibrium structure (Δ[∠H–N–C] = +20.4°, Fig. 2). While the 9a' orbital is formally non-bonding, we suggest the origin of this distortion to be accounted for by the small contribution of the in-plane H 1s to the MO [8]. As the 2a'' orbital is almost entirely non-bonding, and removal of one electron from this orbital causes very little change in geometric structure [8], we expect the band from pure $2a''^{-2}$ ionization to be narrow, without extended vibrational structure. Calculations at the CASSCF/MRCI level of theory predict the vertical double ionization to the 1A' ($2a''^{-2}$) state to be only 0.2 eV greater than the adiabatic double ionization to the same state (Fig. 2), suggestive of minimal structural distortion. The second major peak in the 100 eV double ionization spectrum at 38.5 eV, with a 1.3 eV half-width, is marginally narrower than the first. This is a normal width for a vibrationally extended single band, and is almost at the apparatus-limited resolution of 1.2 eV in 100 eV double photoionization (60 eV electron energy). The first band in the 100 eV double ionization spectra is clearly composite, most probably containing all four possible states from $9a'^{-2}$, $2a''^{-1}9a'^{-1}$ and $2a''^{-2}$ ionizations. Similarly, the second major band at 38.5 eV must involve ionization from the 8a' and 1a'' orbitals with closely spaced binding energies, as confirmed by the calculated singlet state energies [5] and by sharpness of this band in the Auger spectra. It seems most probable that the long-lived HNCO$^{2+}$ ions seen in the mass spectrum at 200 eV electron energy by Wang et al. [12] are in the ground $^3$A'' state created by $2a''^{-1}9a'^{-1}$ ionization. This is essentially confirmed by the ion-coincident double ionization spectra shown in Fig. 3.

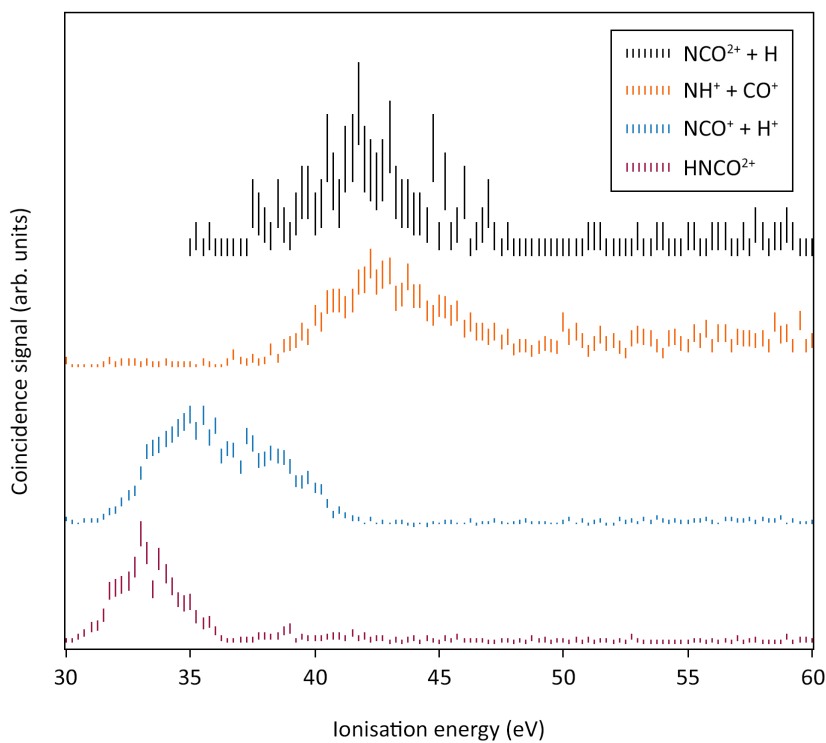

Figure 3. Double ionization spectra coincident with the ions and ion pairs indicated, measured at 100 eV photon energy. The four spectra are normalised to the same maximum amplitude for display, because the true relative amplitudes were not directly measurable (see text). The energy resolution is estimated as about 3 eV at ionization energies near 30 eV. Error bars show only the statistical uncertainty of the coincidence counts included.

To obtain the spectra for the ionic products (Fig. 3) free from contamination by other ions, it was necessary to strictly limit the ion flight-time ranges attributed to each, arbitrarily cutting down the total coincidence signals. For the ion pair products, the signals are also significantly weakened by loss of ions with initial sideways momentum, which hit the spectrometer walls or electrodes and so do not reach the detector. When the estimated effect of these losses is included, the full intensities of the channels are estimated as $HNCO^{2+}$, 78 %, $H^+ + NCO^+$, 52% and $NCO^{2+}$, 12 % relative to $NH^+ + CO^+$ as 100 %. These estimated intensities are in similar proportions as the cross-sections given by Wang et al. [12] for the same channels under 200 eV electron impact. The spectrum coincident with undissociated $HNCO^{2+}$ is consistent, in view of the estimated 3 eV energy resolution, with stability of this ion in a narrow energy range near or at the double ionization onset. The appearance of $NCO^{2+}$ at 38±1 eV, combined with the dissociation energy of HNCO by H—NCO bond cleavage, implies a double ionization energy of NCO as 33±1 eV, essentially the same as that of HNCO itself. It also means that this dissociation takes place without any substantial kinetic energy release and so no reverse activation energy in the pathway. For the $H^+ + NCO^+$ pair, the observed appearance energy near 33 eV implies kinetic energy release of about 5 eV, which is entirely normal for such a charge separation forming ground state products. The appearance of the $NH^+ + CO^+$ pair, on the other hand, is delayed to about 38 eV, nearly 7 eV above its thermodynamic threshold. This is probably a "kinetic shift", meaning that no substantial production of the pair can occur in competition with $H^+$ + $NCO^+$ until the rate of the new reaction outcompetes that of the established lower-energy process.

Because of spectral congestion, kinetic energy releases in the ion-pair production reactions could not be measured directly in these experiments. We return to a general discussion of the dissociation processes of multiply charged HNCO ions in a later section.

In view of the stability and abundance of an $[H,N,C,O]^{++}$ doubly charged ion, it is pertinent to ask if $HNCO^{++}$ or another isotopic form is the most stable. Earlier work undertaken by Morokuma gives a relative energy ordering for the four most stable isomers of [H,N,C,O] (HNCO < HOCN < HCNO < HONC) for both the neutral and the singly-ionized states, calculated at the B3LYP/6-311G(*d,p*) level of theory [9.20]. In new calculations at the same level of theory, as expected the triplet state is favoured over the singlet for each of the doubly-ionized isomers. The ordering of isomer energies is unchanged: HNCO < HOCN < HCNO < HOCN. We do note, however, that a substantial increase in static correlation in the doubly-ionized species may result in poor performance of B3LYP for these systems. For example, for HNCO, the singlet-triplet splitting was calculated to be 1.58 eV, compared with 1.32 eV calculated at the CASSCF/MRCI level of theory.

**Core-valence (CV) spectra**

In a second form of double ionization one electron is removed from a valence orbital and another is removed from a core orbital, here the 1s orbital of one of the C, N or O atoms. Because the core orbitals are remote in space and energy from the valence orbitals they hardly affect the molecular bonding, apart from the electrostatic influence of the localised core charges. As a result, core-valence double ionization spectra are closely related to the photoelectron spectra from single valence electron ionization. The three core-valence double ionization spectra of HNCO in Fig. 4 illustrate this relationship.

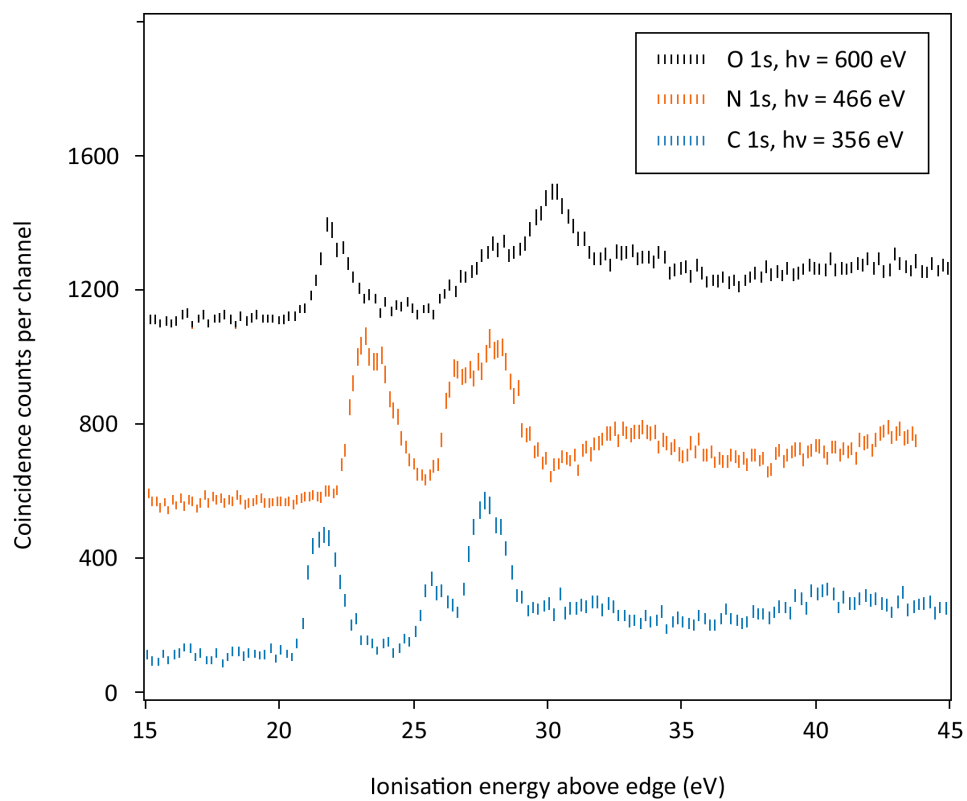

Figure 4.  Core-valence spectra of HNCO above each of the edges, at the photon energies shown.

Each of the core-valence spectra in Fig. 4 has four main bands with spacings between them similar to the spacings of bands in the photoelectron spectrum.  The comparison is further illustrated in Fig. 5, where the core-valence spectrum above the C1s edge is contrasted with a photoelectron spectrum measured in the same apparatus at 100 eV photon energy, where the instrumental resolution is about the same.

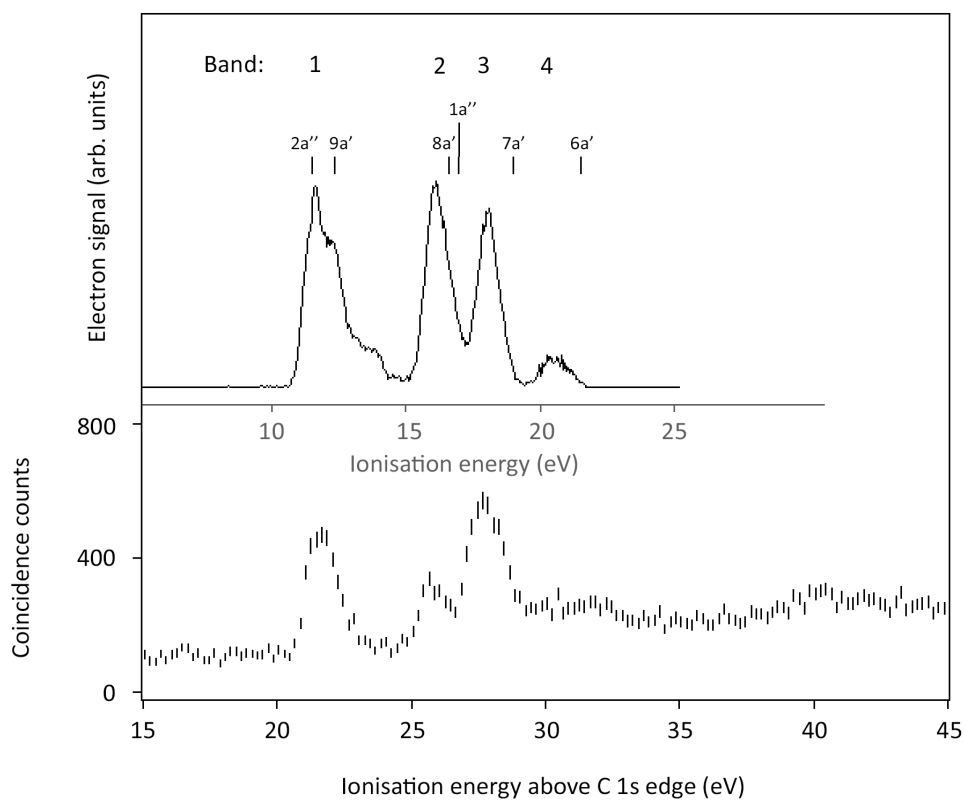

Figure 5. The $C1s^{-1}V^{-1}$ spectrum of HNCO contrasted with a photoelectron spectrum from single ionization at 100 eV photon energy. Calculated binding energies of the first six orbitals [5,8], shifted to fit to the first photoelectron band, are shown above, with identification of the four composite bands discussed in the text.

In Fig. 5, and in comparisons between the other two core-valence spectra and the photoelectron spectrum, the numbers and spacings of bands in the two sorts of spectra are similar. This indicates first of all that the singlet-triplet splittings of the core-valence states are relatively small, as also observed in the analysis of some other core-valence spectra [26,27]. To understand the differing shifts and relative intensities seen in Fig. 3 we propose a very simple model, which should apply to molecules about as big as HNCO or bigger. It is less likely to work for small molecules such as diatomics, where spin-spin and other complications may loom larger [24,25]. For such molecules we express the core-valence ionization energy $E_j$ from each orbital j above the relevant edge as

$$IE_j - E_{edge} = IP_j + e^2/r_{12}$$

where $IP_j$ is the orbital binding energy in the neutral molecule taken from the photoelectron spectrum and $e^2/r_{12}$ is a notional Coulomb repulsion energy between the charge in the delocalised molecular orbital and the localised core charge. Since the orbital ionization energies are known from the photoelectron spectra [8, 17] the measured ionization energies of each band above the related edge gives the notional Coulomb energy and apparent inter-charge distance $r_{12}$. The general suitability of this model is clearly demonstrated by systematic changes in the notional Coulomb energy as a function of the size of the molecules. In molecules with just one heavy atom such as HCl, $H_2O$ or $NH_3$ it is between 12 and 16 eV. In $CF_4$ it is about 8 eV, in $SF_6$ and $C_6H_6$ it is about 5 eV and in $C_{60}$ it falls to about 2 eV [23]. The apparent inter-charge distances $r_{12}$ from the Coulomb energies correspond in every case to the approximate dimensions of the molecules. Within a single molecule, explanation of differences in the Coulomb shifts in ionization from different valence orbitals calls for an extension of the model. To approach this, we consider the characters of the molecular orbitals, particularly the spatial distribution of their charge densities.

Because the in-plane and out-of-plane orbitals in HNCO corresponding to each of $\pi_g$ and $\pi_u$ in $CO_2$ are mostly not resolved, we clump these together and discuss the spectra in terms of four composite bands, 1 – 4, as shown Fig. 5 and listed in Table 1.

A first observation is that in the CV spectrum of HNCO above the C edge, Fig. 5, the spacing of the four bands closely matches the spacing of the bands in the photoelectron spectrum, whereas the spacings are markedly different when the core-hole is on N or O (Fig. 4). The apparent Coulomb energies and intercharge distances relevant to each band can be derived from comparison of the core-valence spectra with the photoelectron spectrum [8, 21]. Apparent intercharge distances for all four bands in the carbon-edge CV spectrum are about 1.4 Å, which is a bit larger than the C—N and C—O bond lengths (1.21 and 1.17 Å) in the neutral molecule [6-8,26]. Some bands in the N- and O- edge CV spectra exhibit larger apparent Coulomb energies and shorter intercharge distances (Bands 1 and 4 in the N-edge CV spectrum, bands 2, 3 and 4 in the O-edge CV spectrum). No band exhibits a longer apparent intercharge distance, for example none is near the overall length of the molecule of 2.5 Å. These characteristics must be related to the forms of the molecular orbitals, and as a rough guide to these, we can use the atomic orbital coefficients tabulated by Holzmeier et al. [5] for the neutral ground state molecule.

The summed squares of the orbital coefficients for the six orbitals that contribute to bands 1 to 4, Table 2, have characteristics that relate clearly to the CV spectra. First, no orbital (or $\pi_g$/ $\pi_u$ pair) has any strong concentration on the C atom. This is consistent with the close match between the CV spectrum above the C1s edge and the photoelectron spectrum. The greatest concentration of all (**bold** in Table 2) is of the $\pi_g$ pair (band 1) on the N atom, which explains the large apparent Coulomb energy (11.5 eV, $r_{12}$ = 1.25 Å) shown by this band when the core charge is on N. The next notable concentrations are of the orbitals for bands 2 and 3 on the O atom, again explaining those bands' significant shifts to higher ionization energy when the core charge is on O. For the N-edge CV spectrum, the orbitals corresponding to bands 2 and 3 have no strong concentration on N, while the orbital for band 4 is concentrated there, compatible with its larger apparent Coulomb energy. Overall, the qualitative agreements found in this way and demonstrated in the comparison between Tables 1 and 2 clearly show that the basic physics underlying the formation of the CV spectra is expressed in the Coulomb model.

This simple molecular orbital and Coulomb model works in other cases too. In $CO_2$, whose outermost ($\pi_g$) orbital is located only on the O atoms, the first band in the CV spectra is shifted to higher energy when the core hole is on an O atom than when it is on the C atom. In the CV spectrum of CF$_4$ with a hole in C1s, the bands from orbitals with C—F bonding character are shifted to higher energy than those with pure F lone-pair character. Similarly for acetaldehyde, whose HOMO is located strongly on the O atom, the lowest energy CV band is at considerably higher energy when the core hole is on the O atom than when it is on either of the C atoms.

**Triple ionization**

HNCO can be triply ionized directly at photon energies below and above all the inner shells, by double Auger decay from the three 1s$^{-1}$ hole states, or by single Auger decay from the doubly ionized CV states discussed above. In practice, the triple ionization cross-section at photon energies below the inner shells is too small to give a significant signal above background in the present experiments. Of the other possibilities, only Auger decay from the C1s hole state and from the associated CV states, giving Auger electron energies of about 200 to 250 eV, offers useful electron energy resolution with the present apparatus. Three spectra are presented in Fig. 6.

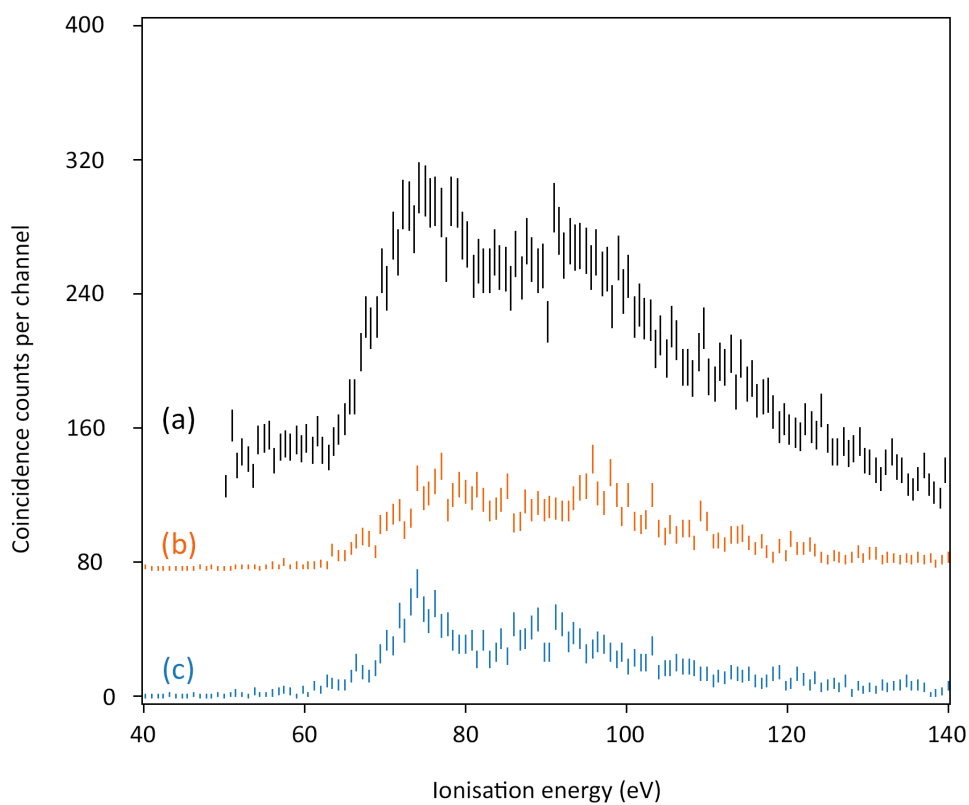

Figure 6. Triple ionization spectra of HNCO. (a): from double Auger decay of the C1s hole state at 296.0 eV using 316 eV photon energy; (b): from single Auger decay of the lowest energy core-valence doubly-ionized state at ca 318 eV (band 1); (c): from single Auger decay of the core-valence state at about 324 eV (Band 3)). Photon energy for the CV spectra was 356 eV.

The three triple ionization spectra in Fig. 6 are similar in form, with onset at about 65 eV and two broad peaks, the first near 75 eV and a second one 20 or 25 eV higher in energy.

Because of the likelihood of nuclear motion in an intermediate state (core-hole or valence hole(s) for spectrum (a), core valence states for (b) and (c)) the onset energies cannot be assumed to represent pure vertical transitions. However, a calculation at CASSCF/MRCI gives the vertical triple ionization energy as 64.8 eV, agreeing with the onsets of all three forms of the triple ionization spectra in Fig. 6 at 65±1 eV. This is considerably lower than the triple ionization energy of $CO_2$, previously determined by a similar method as 74±0.5 eV [27]. The lack of fine detail in the triple ionization spectra is caused partly by the instrumental resolution (ca. 4 eV) but also by the expected congestion of electronic states. The removal of three electrons from the six outer valence orbitals can give rise to 55 electronic states, 35 doublets and 20 quartets, within an estimated 40 eV energy range. All are likely to be dissociative, giving broad Franck-Condon envelopes.

**HNCO di- and tri-cation dissociation dynamics**

The pathways followed in fragmentation of nascent $HNCO^{2+}$ and $HNCO^{3+}$ and their partial cross-sections were determined by Wang et al. [11] in their covariance analysis of electron-impact-induced dissociative ionization. The present results from photon impact at 100 eV photon energy confirm that the states created by 200 eV electron impact are much the same as those created in 100 eV photon impact. Unfortunately neither we nor Wang et al. have yet measured the kinetic energy releases in the different channels, so we cannot relate different channels directly to different elements of the spectra. It is nevertheless useful to compare the thermodynamic thresholds for all the different channels with the relative partial cross-sections and the spectra, with the usual magnitudes of kinetic energy releases in mind. The thresholds for ground-state products determined from the tabulation of Lias et al. [28] are listed in Table 3 for the 7 most abundant dication dissociation pathways and six trication dissociation channels listed by Wang et al. [12]. The appearance energies are estimated for direct dissociative double ionization, by adding an assumed minimum kinetic energy release (KER) based on a probable intercharge distance. Pathways involving autoionization of superexcited neutral fragments, which could have much lower appearance energies, are excluded.

From the presence of $HNCO^{2+}$ ions in the mass spectrum and its appearance near the lowest double ionization energy as shown in in Fig. 3, we conclude that this ion is stable or metastable in its ground state. There must be a barrier to dissociation towards $H^+ + NCO^+$, the only competing pathway, whose thermodynamic limit is at lower energy. Even for the proton the lifetime towards tunnelling through the barrier is evidently long enough (≈μs) for mass-spectroscopic observation and may be much longer. A calculation of the potential energy surfaces involved would be very interesting. We note that both of the observed two-body dissociations of $HNCO^{++}$ are spin- and symmetry-allowed, $H^+(^1\Sigma^+) + NCO^+(^3\Sigma^-)$ and $NH^+(^2\Pi) + CO^+(^2\Pi)$ and correspond to simple bond breakages. The dissociation to $H + NCO^{2+}$ is also spin-allowed and we suspect that the ground state potential energy surface of $HNCO^{2+}$ correlates to this limit. The barrier to charge separating dissociation may arise from an avoided crossing with the surface leading to $H^+ + NCO^+$. No strong signal due to any intra-molecular rearrangement is observed, in contrast to the situation in the singly charged ion [8], perhaps because decays of the doubly charged ions are too rapid. Three-body fragmentations all start at higher energies roughly within the intense bands above 40 eV ionization energy.

In contrast to the double ionization, triple ionization fragmentation thresholds are all well below the levels populated in the spectra of Fig. 6. For the kinetic energy release we have adopted a single value of 10 eV. Even with this energy added, the estimated appearance energies are all below or within the lower energy band of the triple ionization spectrum, so the absence of a detected $HNCO^{3+}$ ion [11] is unsurprising. Three-body fragmentations occur at the lowest energies, but four-body decays with complete atomization of the molecule are possible for the majority of the levels seen to be populated in Fig. 6. We cannot assume that the pattern of population in triple ionization by photons or electrons with energies below the inner-shell threshold will match the observed spectra of Fig. 6. But directly populated triply ionized states are unlikely to lie lower in ionization energy than the states behind the spectra shown in Fig. 6; instead they may lie higher in energy, as intermediate core-hole states, where partial dissociation can occur, are not involved.

**Conclusions**

The valence double ionization spectrum of HNCO and spectra coincident with its dissociation products, compared with the thermodynamic thresholds for two body dissociations suggest that in astrophysical environments double ionization by cosmic ray or EUV (extreme ultraviolet) impact will completely destroy the molecule. Core-valence double ionization, while probably irrelevant to astrophysics, provides an interesting testing ground for simple physical theory. We show that a very simple Coulomb model explains the main features of the core-valence spectra of HNCO, and by implication those of many other compounds. A first triple ionization spectrum is also reported.

**Acknowledgements**

This work has been financially supported by the Swedish Research Council (VR) and the Knut and Alice Wallenberg Foundation, Sweden. We thank the Helmholtz Zentrum Berlin for the allocation of synchrotron radiation beam time and the staff of BESSY-II for particularly smooth running of the storage ring during the single-bunch runtime. The research leading to these results has received funding from the European Union's Horizon 2020 research and innovation programme under grant agreement No 730872.

**Table 1.** HNCO core-valence band Coulomb energies. Larger values **bold**.

| Band (IP/eV): | | 1 (12) | 2 (15.5) | 3 (17.5) | 4 (21) |
|---|---|---|---|---|---|
| Atom | Edge/eV | Coulomb energies (sh = shoulder) | | | |
| C | 296.0 | 9.6 | 10.2 | 10.2 | 10.7 |
| N | 406.3 | **11.4** | **11.3**(sh) | 10.6 | **12.2** |
| O | 539.9 | 10.0 | **11.8** | **12.5** | **12.5** |

**Table 2.** MO relative atom densities in HNCO from Ref. [5]. Largest density in each orbital or combination in **bold**.

| Band: | 1 | 2 | 3 | 4 |
|---|---|---|---|---|
| C | 0.05 | 0.42 | 0.14 | 0.13 |
| N | **1.26** | 0.23 | 0.06 | **0.44** |
| O | 0.73 | **0.79** | **0.84** | 0.10 |

**Table 3.** Thresholds for $HNCO^{2+}$ and $HNCO^{3+}$ charge-separating dissociations

*Channels listed in decreasing order of partial cross-section in 200 eV electron impact [11].*

| Channel | Thermodynamic limit (eV) | Appearance energy (eV) | |
|---|---|---|---|
| | | Estimated | Observed |
| $NH^+ + CO^+$ | 31.3 | 35 | 38±1 |
| $H^+ + N^+ + CO$ | 35.2 | 40 | |
| $H^+ + O^+ + CN$ | 37.5 | 42 | |
| $H^+ + NCO^+$ | 30.2 | 33.5 | 33±0.5 |
| $H^+ + C^+ + NO$ | 36.6 | 41 | |
| $C^+ + N^+ + OH$ | 39.5 | 44 | |
| $N^+ + CO^+ + H$ | 35.6 | 41 | |
| | | | |
| $H^+ + N^+ + O^+ + C$ | 60.0 | 70 | |
| $H^+ + C^+ + O^+ + N$ | 56.7 | 67 | |
| $H^+ + C^+ + N^+ + O$ | 57.5 | 68 | |
| $H^+ + N^+ + CO^+$ | 49.2 | 59 | |
| $C^+ + N^+ + O^+ + H$ | 57.6 | 68 | |
| $H^+ + O^+ + CN^+$ | 51.6 | 62 | |